# Tangential force, Frictional Torque and Heating Rate of a Small Neutral Rotating Particle Moving through the Equilibrium Background Radiation

G.V. Dedkov[1] and A.A. Kyasov

Nanoscale Physics Group, Kabardino-Balkarian State University, Nalchik, 360004, Russia

For the first time, based on the fluctuation-electromagnetic theory, we have calculated the drug force, the radiation heat flux and the frictional torque on a small rotating particle moving at a relativistic velocity through the equilibrium background radiation (photon gas). The particle and background radiation are characterized by different temperatures corresponding to the local thermodynamic equilibrium in their own reference frames.

## 1. Introduction

Interaction of a small neutral polarizable body with vacuum electromagnetic fields is the long-standing classical problem of the fluctuation and quantum electrodynamics [1]. Of particular interest is the case when the particle moves, rotates or simultaneously rotates and moves with relativistic velocity. The first dynamical problem of such a kind has been solved by Mkrtchian et. al. [2], relating to the case of frictional drug acting on a nonrelativistic particle moving through an equilibrium photon gas. This photon gas (heat bath) is formed in an oven or it may represent a cosmic microwave background. It is worth noting that the well-known formula [3] for the radiation force exerted by the photon gas on a relativistic sphere is not adequate in the case under consideration, namely $R << \lambda_W$, with $R$ and $\lambda_W$ being the particle radius and the characteristic wave-length of thermal radiation. The problems of relativistic frictional drug and particle-vacuum radiation (for small particles $R << \lambda_W$) were first examined by us in [4,5]. The results [2] and [4,5] have been also confirmed by other authors [6,7], using a fully covariant calculation method. Quite recently, we and other authors [8-10] have calculated frictional torque and thermal radiation intensity on a particle rotating in vacuum [8], near the surface [9,10], and for two rotating particles in vacuum [11]. The purpose of this work is to consider the more general problem relating to the particle and the background electromagnetic radiation, assuming the particle to be simultaneously moving and rotating. We obtain the general theoretical expressions for the tangential frictional force, frictional torque and the rate of thermal radiation on the particle that incorporate the effects of uniform motion and rotation. All previous results simply follow from the obtained formulas.

---
[1] Corresponding author e-mail: gv_dedkov@mail.ru

## 2. Theory

We will use the same calculation method as that in [4,5,9], considering a small spherical particle of radius $R$ with the dipole electric polarizability $\alpha(\omega)$ and temperature $T_1$, moving with the velocity $\mathbf{V} = (V,0,0)$ and simultaneously rotating with the angular velocity $\mathbf{\Omega} = (\Omega,0,0)$ (Fig. 1a) or $\mathbf{\Omega} = (0,0,\Omega)$ (Fig. 1b) in the space filled by the equilibrium electromagnetic radiation (photonic gas) of temperature $T_2$.

Therefore, the angular velocity $\Omega$ is defined in the frame $\Sigma'$ which moves with the velocity $\mathbf{V}$ relative to the system $\Sigma$ related to the equilibrium radiation. The values $\alpha(\omega)$ and $T_1$ are defined in the particle rest frame $\Sigma''$ rotating with the angular velocity $\Omega$ relative to the frame $\Sigma'$.

Assuming that all the physical quantities are defined in $\Sigma$, the starting equations for the tangential force $F_x$, the frictional torque $M_x$ and the particle heating rate $\dot{Q}$ are given by

$$F_x = \left\langle \nabla_x \left( \mathbf{d}^{sp} \cdot \mathbf{E}^{ind} + \mathbf{m}^{sp} \cdot \mathbf{B}^{ind} \right) \right\rangle + \left\langle \nabla_x \left( \mathbf{d}^{ind} \cdot \mathbf{E}^{sp} + \mathbf{m}^{ind} \cdot \mathbf{B}^{sp} \right) \right\rangle \equiv F_x^{(1)} + F_x^{(2)} \tag{1}$$

$$\dot{Q} = \left\langle \dot{\mathbf{d}}^{sp} \cdot \mathbf{E}^{ind} + \dot{\mathbf{m}}^{sp} \cdot \mathbf{B}^{ind} \right\rangle + \left\langle \dot{\mathbf{d}}^{ind} \cdot \mathbf{E}^{sp} + \dot{\mathbf{m}}^{ind} \cdot \mathbf{B}^{sp} \right\rangle \equiv \dot{Q}^{(1)} + \dot{Q}^{(2)} \tag{2}$$

$$M_x = \left\langle \mathbf{d}^{sp} \times \mathbf{E}^{ind} + \mathbf{m}^{sp} \times \mathbf{B}^{ind} \right\rangle_x + \left\langle \mathbf{d}^{ind} \times \mathbf{E}^{sp} + \mathbf{m}^{ind} \times \mathbf{B}^{sp} \right\rangle_x \equiv M_x^{(1)} + M_x^{(2)} \tag{3}$$

In Eqs. (1)-(3), the superscripts "sp", "ind" denote the spontaneous and induced components of the fluctuating dipole electric and magnetic moments **d, m,** and electromagnetic fields **E, B**, the points above $Q, \mathbf{d}, \mathbf{m}$ denote the time differentiation, while the angular brackets denote complete quantum and statistical averaging. It is worth noting that within the relativistic statement of the problem, even the particle with zero magnetic polarizability in its rest frame $\Sigma''$ has the fluctuating magnetic moment in $\Sigma$.

For definiteness, let us consider configuration 1 (Fig. 1a). In subsequent calculations, one should take into account that in the presence of rotation the fluctuation-dissipation theorem (FDT) for the spontaneous fluctuating dipole moment of the particle in frame $\Sigma'$ takes the form

$$\left\langle d^{sp}{}'_x(\omega')d^{sp}{}'_x(\omega)\right\rangle = 2\pi\hbar\delta(\omega+\omega')\alpha''(\omega)\coth\frac{\hbar\omega}{2k_BT_1} \qquad (4)$$

$$\left\langle d^{sp}{}'_y(\omega')d^{sp}{}'_y(\omega)\right\rangle = \left\langle d^{sp}{}'_z(\omega')d^{sp}{}'_z(\omega)\right\rangle = \frac{1}{2}2\pi\hbar\delta(\omega+\omega')\cdot$$
$$\cdot\left[\alpha''(\omega_+)\coth\frac{\hbar\omega_+}{2k_BT_1}+\alpha''(\omega_-)\coth\frac{\hbar\omega_-}{2k_BT_1}\right] \qquad (5)$$

$$\left\langle d^{sp}{}'_y(\omega')d^{sp}{}'_z(\omega)\right\rangle = -\left\langle d^{sp}{}'_z(\omega')d^{sp}{}'_y(\omega)\right\rangle = -\frac{i}{2}2\pi\hbar\delta(\omega+\omega')\cdot$$
$$\cdot\left[\alpha''(\omega_+)\coth\frac{\hbar\omega_+}{2k_BT_1}-\alpha''(\omega_-)\coth\frac{\hbar\omega_-}{2k_BT_1}\right] \qquad (6)$$

where $\omega_\pm = \omega\pm\Omega$.

Moreover, in the case of rotation, relationships between the induced dipole and magnetic moments of rotating particle and the spontaneous fluctuating electromagnetic field of the equilibrium background radiation in frame $\Sigma$ take the form

$$d^{in}{}_x(t) = \frac{1}{\gamma}\int\frac{d\omega d^3k}{(2\pi)^4}\alpha(\gamma(\omega-k_xV))E^{sp}{}_x(\omega,\mathbf{k})\exp(-i(\omega-k_xV)t) \qquad (7)$$

$$d^{in}{}_y(t) = \gamma\int\frac{d\omega d^3k}{(2\pi)^4}\exp(-i(\omega-k_xV)t)\cdot$$
$$\cdot\frac{1}{2}\{\alpha(\gamma(\omega-k_xV)+\Omega)\cdot[(E^{sp}{}_y(\omega,\mathbf{k})+iE^{sp}{}_z(\omega,\mathbf{k}))-\beta(B^{sp}{}_z(\omega,\mathbf{k})-iB^{sp}{}_y(\omega,\mathbf{k}))]+ \qquad (8)$$
$$+\alpha(\gamma(\omega-k_xV)-\Omega)[(E^{sp}{}_y(\omega,\mathbf{k})-iE^{sp}{}_z(\omega,\mathbf{k}))-\beta(B^{sp}{}_z(\omega,\mathbf{k})+iB^{sp}{}_y(\omega,\mathbf{k}))]\}$$
.

$$d^{in}{}_z(t) = \gamma\int\frac{d\omega d^3k}{(2\pi)^4}\exp(-i(\omega-k_xV)t)\cdot$$
$$\cdot\frac{1}{2}\{\alpha(\gamma(\omega-k_xV)+\Omega)\cdot[(E^{sp}{}_z(\omega,\mathbf{k})-iE^{sp}{}_y(\omega,\mathbf{k}))+\beta(B^{sp}{}_y(\omega,\mathbf{k})+iB^{sp}{}_z(\omega,\mathbf{k}))]+ \qquad (9)$$
$$+\alpha(\gamma(\omega-k_xV)-\Omega)[(E^{sp}{}_z(\omega,\mathbf{k})+iE^{sp}{}_y(\omega,\mathbf{k}))+\beta(B^{sp}{}_y(\omega,\mathbf{k})-iB^{sp}{}_z(\omega,\mathbf{k}))]\}$$
.

$$m^{in}{}_x(t) = 0, m^{in}{}_y(t) = \beta d^{in}{}_z(t), m^{in}{}_z(t) = -\beta d^{in}{}_y(t) \qquad (10)$$

where $\beta = V/c, \gamma = (1-\beta^2)^{-1/2}$.

In the case of configuration 2, the correlators of dipole moments are obtained from (4)—(6) by a cyclic permutation $x \to y \to z \to x$, while induced dipole moments take the form

$$d^{in}_x(t) = \int \frac{d\omega d^3k}{(2\pi)^4} \exp(-i(\omega - k_x V)t) \cdot$$
$$\cdot \left[\gamma^{-1}\alpha_1(\omega,\Omega)E^{sp}_x(\omega,\mathbf{k}) + i\alpha_2(\omega,\Omega)\left(E^{sp}_y(\omega,\mathbf{k}) - \beta B^{sp}_z(\omega,\mathbf{k})\right)\right] \quad (11)$$

$$d^{in}_y(t) = \int \frac{d\omega d^3k}{(2\pi)^4} \exp(-i(\omega - k_x V)t) \cdot$$
$$\cdot \left[-i\alpha_2(\omega,\Omega)E^{sp}_x(\omega,\mathbf{k}) + \gamma\alpha_1(\omega,\Omega)\left(E^{sp}_y(\omega,\mathbf{k}) - \beta B^{sp}_z(\omega,\mathbf{k})\right)\right] \quad (12)$$

$$d^{in}_z(t) = \gamma \int \frac{d\omega d^3k}{(2\pi)^4} \exp(-i(\omega - k_x V)t)\alpha(\gamma(\omega - k_x V))\left[E^{sp}_z(\omega,\mathbf{k}) + \beta B^{sp}_y(\omega,\mathbf{k})\right] \quad (13)$$

$$\alpha_{1,2}(\omega,\Omega) = \frac{1}{2}\left(\alpha(\gamma(\omega - k_x V) + \Omega) \pm \alpha(\gamma(\omega - k_x V) - \Omega)\right) \quad (14)$$

The corresponding magnetic moments are given by Eqs. (10) combined with (11)—(13). Using (4)—(10), the calculations in (1)-(3) are performed very similar to [5] and [12].

## 3. Results

The obtained final expressions for the tangential force $F_x$, the heating rate $\dot{Q}$ and the frictional torque $M_x$ ($M_z$) on a particle are given by ($\beta = V/c$, $\gamma = (1-\beta^2)^{-1/2}$)

a) configuration 1 ($\mathbf{\Omega} = (\Omega,0,0)$)

$$F_x = -\frac{\hbar\gamma}{4\pi c^4} \int_{-\infty}^{+\infty} d\omega \omega^4 \int_{-1}^{1} dx\, x \cdot$$
$$\left\{(1-x^2)(1-\beta^2)\alpha''(\gamma\omega(1+\beta x)) \cdot \left[\coth\frac{\hbar\omega}{2k_B T_2} - \coth\frac{\hbar(\gamma\omega(1+\beta x))}{2k_B T_1}\right] + \right.$$
$$\left. + \left[(1+x^2)(1+\beta^2) + 4\beta x\right]\alpha''(\gamma\omega(1+\beta x) + \Omega) \cdot \left[\coth\frac{\hbar\omega}{2k_B T_2} - \coth\frac{\hbar(\gamma\omega(1+\beta x) + \Omega)}{2k_B T_1}\right]\right\} \quad (15)$$

$$dQ/dt = \frac{\hbar \gamma}{4\pi c^3} \int_{-\infty}^{+\infty} d\omega \omega^4 \int_{-1}^{1} dx (1+\beta x) \cdot$$

$$\left\{ (1-x^2)(1-\beta^2)\alpha''(\gamma\omega(1+\beta x)) \cdot \left[ \coth\frac{\hbar\omega}{2k_B T_2} - \coth\frac{\hbar(\gamma\omega(1+\beta x))}{2k_B T_1} \right] + \right.$$

$$\left. + \left[ (1+x^2)(1+\beta^2) + 4\beta x \right] \alpha''(\gamma\omega(1+\beta x)+\Omega) \cdot \left[ \coth\frac{\hbar\omega}{2k_B T_2} - \coth\frac{\hbar(\gamma\omega(1+\beta x)+\Omega)}{2k_B T_1} \right] \right\} \quad (16)$$

$$M_x = -\frac{\hbar\gamma}{4\pi c^3} \int_{-\infty}^{+\infty} d\omega \omega^3 \int_{-1}^{1} dx \left[ (1+x^2)(1+\beta^2) + 4\beta x \right] \alpha''(\gamma\omega(1+\beta x)+\Omega) \cdot$$

$$\cdot \left[ \coth\frac{\hbar\omega}{2k_B T_2} - \coth\frac{\hbar(\gamma\omega(1+\beta x)+\Omega)}{2k_B T_1} \right] \quad (17)$$

b) configuration 2 ($\mathbf{\Omega} = (0,0,\Omega)$)

$$F_x = -\frac{\hbar\gamma}{8\pi c^4} \int_{-\infty}^{+\infty} d\omega \omega^4 \int_{-1}^{1} dx\, x \cdot$$

$$\left\{ \begin{array}{l} \left[ (1+x^2)(1+\beta^2) + 2(1-\beta^2)(1-x^2) + 4\beta x \right] \alpha''(\gamma\omega(1+\beta x)+\Omega) \cdot \\ \cdot \left[ \coth\frac{\hbar\omega}{2k_B T_2} - \coth\frac{\hbar(\gamma\omega(1+\beta x)+\Omega)}{2k_B T_1} \right] + \left[ (1+x^2)(1+\beta^2) + 4\beta x \right] \alpha''(\gamma\omega(1+\beta x)) \\ \cdot \left[ \coth\frac{\hbar\omega}{2k_B T_2} - \coth\frac{\hbar(\gamma\omega(1+\beta x))}{2k_B T_1} \right] \end{array} \right\} \quad (18)$$

$$dQ/dt = \frac{\hbar\gamma}{8\pi c^3} \int_{-\infty}^{+\infty} d\omega \omega^4 \int_{-1}^{1} dx \cdot (1+\beta x)$$

$$\left\{ \begin{array}{l} \left[ (1+x^2)(1+\beta^2) + 2(1-\beta^2)(1-x^2) + 4\beta x \right] \alpha''(\gamma\omega(1+\beta x)+\Omega) \cdot \\ \cdot \left[ \coth\frac{\hbar\omega}{2k_B T_2} - \coth\frac{\hbar(\gamma\omega(1+\beta x)+\Omega)}{2k_B T_1} \right] + \left[ (1+x^2)(1+\beta^2) + 4\beta x \right] \alpha''(\gamma\omega(1+\beta x)) \cdot \\ \cdot \left[ \coth\frac{\hbar\omega}{2k_B T_2} - \coth\frac{\hbar(\gamma\omega(1+\beta x))}{2k_B T_1} \right] \end{array} \right\} \quad (19)$$

$$M_z = -\frac{\hbar}{8\pi c^3} \int_{-\infty}^{+\infty} d\omega \omega^3 \int_{-1}^{1} dx (3-x^2+2\beta x)\alpha''(\gamma\omega(1+\beta x)+\Omega) \cdot$$

$$\cdot \left[ \coth\frac{\hbar\omega}{2k_B T_2} - \coth\frac{\hbar(\gamma\omega(1+\beta x)+\Omega)}{2k_B T_1} \right] \quad (20)$$

As we can see, effects of relativistic motion and rotation form universal combinations in the frequency arguments. The modified frequencies in the polarizability $\alpha(\omega)$ and in the terms involving the particle temperature $T_1$ appear due to the Lorentz and rotation transformations of the electromagnetic field and dipole moments from reference frame of vacuum to the reference frame of the particle. The lack of this dependence in the arguments of cotangent involving vacuum temperature $T_2$ is the characteristic mark of the reference frame at rest. One can also see that the integrands in Eqs. (15),(16) and in Eqs. (18),(19) are completely identical with the replacement $x$ by $(1+\beta x)$. The difference is due to the differentiation over the $x-$coordinate in (1) and over the time $t$ in (2). At $\beta=0$, Eq. (16),(17) and (19),(20) are reduced to the results by Manjavacas et. al. [1] for the particle rotating in vacuum, while at $\Omega=0$ we obtain $M_x=0$ from (17) and $M_z=0$ from (19). In addition, Eqs.(15),(18) are reduced to [4,5]. Finally, one can see that the value of the frictional moment in configuration 1 is a factor $\gamma$ higher than that in configuration 2.

For a particle with the magnetic polarizability $\alpha_m(\omega)$, formulas (15)-(20) remain the same with the replacement $\alpha(\omega) \to \alpha_m(\omega)$, while in general, when the particle has both electric and magnetic moments, we have to take the sum of the polarizabilities.

## Conclusions

We have performed a generalization of the theory of the fluctuation-electromagnetic interaction relating to a small polarizable particle rotating with the angular velocity $\Omega$ and uniformly moving through the equilibrium electromagnetic radiation with relativistic velocity $V$. The particular cases of spinless moving particle and particle with spin at rest follow from the obtained formulas in a simple way.

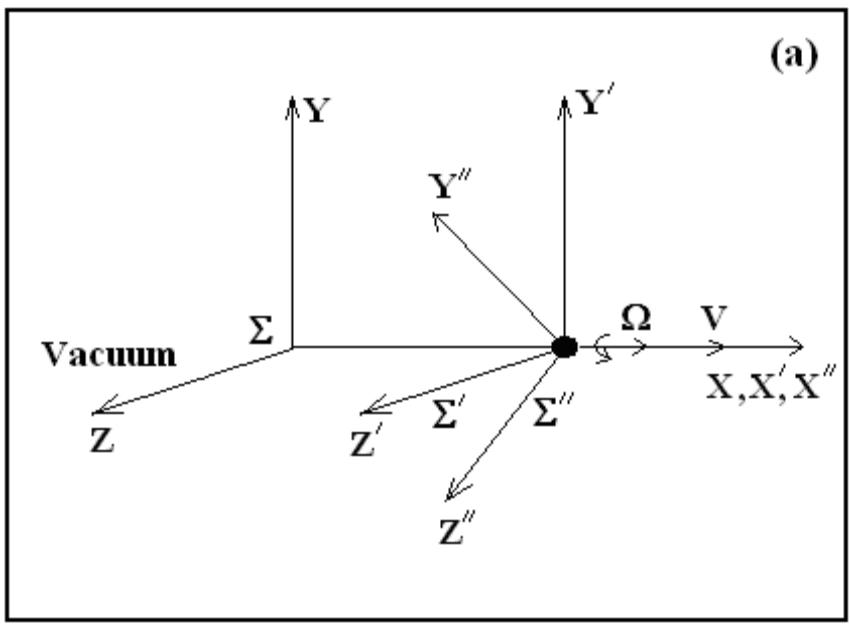

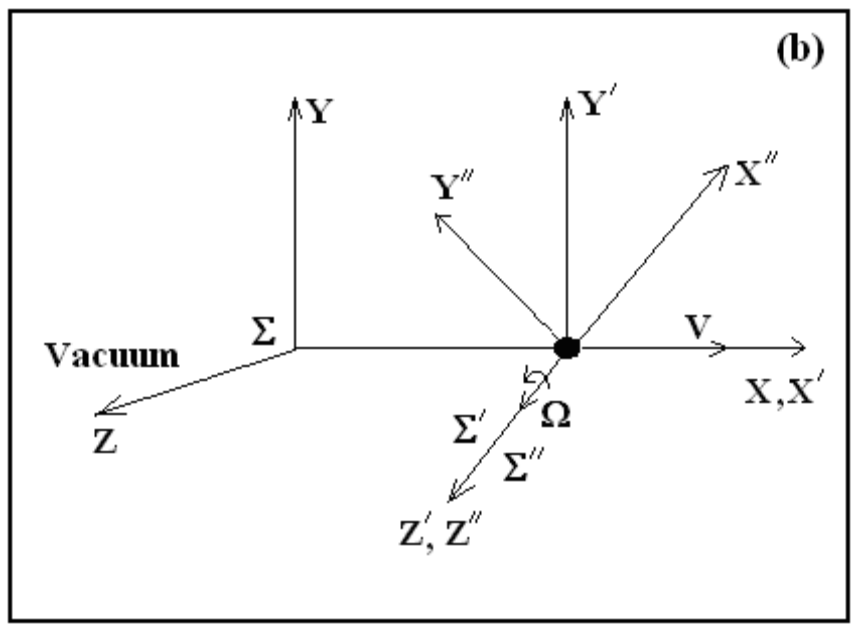

Fig. 1 Geometrical configurations 1 (a) and 2 (b), and coordinate systems